\documentclass{aa}  

\usepackage{graphicx}
\usepackage{txfonts}
\usepackage{hyperref}
\hypersetup{
    colorlinks=true,
    linkcolor=blue,
    citecolor=blue,
    urlcolor=cyan,
    }

\newcommand{\Rlm}[1]{\mathbf{R}^{m_{#1}}_{l_{#1}}}
\newcommand{\Slm}[1]{\mathbf{S}^{m_{#1}}_{l_{#1}}}
\newcommand{\Tlm}[1]{\mathbf{T}^{m_{#1}}_{l_{#1}}}
\newcommand{\Ylm}[1]{Y_{l_{#1}}^{m_{#1}}}
\newcommand{\bnab}{\boldsymbol{\nabla}}

\newcommand{\gradperp}{\bnab_\perp}

\newcommand{\dth}{\partial_\theta}

\newcommand{\dphi}{\partial_\varphi}
\newcommand{\er}{\mathbf{e}_r}
\newcommand{\etheta}{\mathbf{e}_\theta}
\newcommand{\ephi}{\mathbf{e}_\varphi}

\begin{document}

   \title{Magnetochronology of solar-type star dynamos}


   \author{Q. Noraz\,\inst{1,2,3}\thanks{E-mail: quentin.noraz@astro.uio.no}, A. S. Brun\,\inst{1}, 
          \and
          A. Strugarek\,\inst{1}
          }

   \institute{D\'epartement d'Astrophysique/AIM, CEA/IRFU, CNRS/INSU, Univ. Paris-Saclay \& Univ. de Paris, 91191 Gif-sur-Yvette, France\\
             \and
             Rosseland Centre for Solar Physics, University of Oslo, P.O. Box 1029 Blindern, Oslo, NO-0315, Norway\\
             \and
             Institute of Theoretical Astrophysics, University of Oslo, P.O.Box 1029 Blindern, Oslo, NO-0315, Norway.\\
             }

   \date{Received 11 September, 2023; accepted 23 January, 2024}

 
   \abstract
   {}
   {In this study, we analyse the magnetic field properties of a set of 15 global magnetohydrodynamic (MHD) simulations of solar-type star dynamos conducted using the ASH code. Our objective is to enhance our understanding of these properties by comparing theoretical results to current observations, and to finally provide fresh insights into the field.}
   {We analysed the rotational and magnetic properties as a function of various stellar parameters (mass, age, and rotation rate) in a ‘Sun in time’ approach in our extended set of 3D MHD simulations. To facilitate direct comparisons with stellar magnetism observations using various Zeeman-effect techniques, we decomposed the numerical data into vectorial spherical harmonics.}
   {A comparison of the trends we find in our simulations set reveals a promising overall agreement with the observational context of stellar magnetism, enabling us to suggest a plausible scenario for the magneto-rotational evolution of solar-type stars. In particular, we find that the magnetic field may reach a minimum amplitude at a transition value of the Rossby number near unity. This may have important consequences on the long-term evolution of solar-type stars, by impacting the relation between stellar age, rotation, and magnetism. This supports the need for future observational campaigns, especially for stars in the high Rossby number regime.}
   {}


   \maketitle
%

\section{Introduction}
Since the pioneering work of \citet{Skumanich1972}, followed by \citet{Barnes2003}, there has seemed to be a link between the age of a solar-type star and its rotation rate. Older stars usually rotate slower than their younger equivalent; this is the well-known ‘gyrochronology’ proposed by \citet{Barnes2003}. This in turn seems to influence the degree of magnetic activity that a given solar-type star harbours. Young stars are usually much more magnetically active than older ones like the Sun. \citet{Vidotto2014a} propose that this link between stellar age and magnetic activity be called ‘magnetochronology’ (see also \citealt{mathurMagneticActivityEvolution2023}). 

Recently, some authors \citep{VanSaders2016,Hall2021,metcalfe2022} have questioned this link between rotation, magnetic activity, and age, arguing that after a certain age (about the age of the Sun for solar twins, i.e 4.5 Gyr), such a link is broken, with stellar rotation being ‘stalled’. Others have found that it still holds \citep{lorenzo-oliveira2016,lorenzo-oliveira2019,DoNascimento2020}, with possibly only a temporary pause \citep{Curtis2020}. The difficulty comes from the large uncertainty of the age determination and the observational method used. Supporters of the ‘stalling’ scenario are usually basing their analysis on asteroseismically determined ages  and rotation period calibration using Kepler data. So it is important to also consider theoretical aspects when discussing the relation between age, rotation, and magnetic activity levels in the so-called ‘Sun in time’ approach \citep{ayres1997,guinan2002,Ribas2005,gudel2007,Ahuir2020,lorenzo-oliveira2020,Johnstone2021}. For instance, one can turn to numerical simulations of solar-like star dynamos to assess trends between various stellar parameters characterising mass, age, rotation rates, or magnetic states. Since the work of \citet{Durney1977a}, it has been quite clear that the dynamo number $D$, characterising the state of the dynamo, can be directly linked to the Rossby number, $Ro$, of the star such that $D \propto 1/Ro^2$, in classical $\alpha-\omega$ dynamos. The Rossby number is a key non-dimensional number that is widely used in the study of stellar evolution and activity and can be used to bridge observations and numerical simulations \citep{Brun2017,kapyla2023}. Simply put, it allows us to quantify if the turbulence and internal magnetohydrodynamics in rotating stars is strongly influenced (small Rossby numbers) or not (large Rossby numbers) by rotational effects. The most classical definition of the Rossby number is the so-called ‘stellar Rossby number’, and is defined as the ratio between a measure of the convective turnover time, $\tau_c$, at a given depth of the convective envelope and the stellar rotation period, $P_{rot}$; that is, $Ro_s=P_{rot}/\tau_c$ (see \citealt{Landin2010,Brun2017a} for further discussions on the many definitions of this number). When $Ro_s$ is small, rotation influences the convection dynamics, tending towards the so-called magnetostrophic state, when Lorentz and Coriolis forces (and horizontal pressure gradients) balance one another out, for very small Ros values \citep{davidson2014,Augustson2019}. Thanks to an extensive dynamo study published in \citep{Brun2017a,Brun2022} with the ASH code along with a similar study with the Eulag-MHD code \citep{Strugarek2017,Strugarek2018}, we now have a database of more than 30 fully 3D magnetohydrodynamic (MHD) rotating convective dynamos of solar-type stars, spanning several mass and rotation bins; hence, Rossby numbers. In the present paper, we wish to study how the properties of the surface magnetism change as we vary the Rossby number. We are helped by an equivalent systematic observational study of solar-like star magnetism performed by \citet{See2015,See2019,See2019b} (see also \citealt{reiners2022}) in the context of the Bcool consortium \citep{Marsden2014} using Zeeman-Doppler imaging (ZDI) techniques. 
\begin{figure*}
    \sidecaption
        \includegraphics[width=12cm]{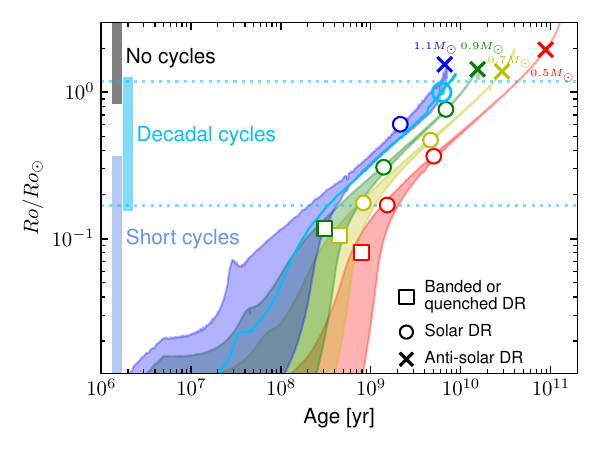}
         \caption{Stellar evolutionary tracks in a Rossby vs age diagram. Shown on the figure are, 1D evolutionary track from \citet{amard2020} as a colour shaded envelope taking into account the initial rotational spread. The particular case of the solar track is shown in cyan colour. Symbols represent the ASH simulations and are changed according to the internal differential rotation profile achieved in the convective envelope of the simulations, either banded/quenched, solar like (fast equator slow poles) and anti-solar (slow equator fast poles). The coloured vertical bars on the left represent the magnetic states either short, decadal and no cycles.}
         \label{fig:0}
\end{figure*}

In Sect. 2, we briefly present the set of 15 ASH simulations published in \citet{Brun2022} that we will analyse further in Sect. 3. We conclude in Sect. 4 by proposing a plausible magneto-rotational scenario over secular timescales for solar-type stars.

\section{Magnetic properties of magnetohydrodynamic models of stellar dynamos}\label{sec:sect2}

\subsection{Brief overview of stellar dynamo simulation database}

In this paper, we focus on the ASH dynamo simulations published in \citep{Brun2017a,Brun2022}.
The properties of the various large-scale flows and magnetic states, as well as their associated energy and non-linear angular momentum transports in purely HD and MHD conditions, have been characterised in detail in these two publications and will not be repeated here. Instead, we wish to focus the analysis in this paper on the global properties of their magnetic field with respect to various stellar parameters.

The simulations represent solar-type stars in a mass range of 0.5 to 1.1 $M_{\odot}$,
with rotation rates of 1/8 to 5 times $\Omega_{\odot}$ at solar metallicity. The effective temperature range considered for the simulations lies approximately between 4030 and 6030 K,  covering mostly G and K-type stars on the main sequence.

In all the simulations, the rotating convective envelope of the stars is modelled from the base of their convection zone up to about 0.97 $R_*$. The range of values covered by the stellar radius is between about 0.44 to 1.23 $R_{\odot}$ and the luminosity ranges from about 0.04 to 1.8 times the solar luminosity, $L_{\odot}$. In the simulations computed with the ASH code, the models also include a stable radiative layer about the same thickness as the convective envelope above, and hence possess a tachocline at the base of their convective zone (in the middle of the computational domain, approximately). The diffusivity profiles (viscous, thermal, and magnetic) are adapted (tapered) such that they maintain an almost constant Reynolds number throughout the simulations. All these simulations develop a genuine multi-scale convective dynamo, and most have been numerically integrated for several decades of physical time over many years.

From a numerical point of view, the ASH code is a semi-implicit pseudo-spectral method of solving the anelastic MHD equations in a frame rotating at $\Omega_*$ \citep{Clune1999,Brun2004}. Each simulation has a significant stratification, the level of which depends on the spectral type considered. The radial density contrast varies from about 40 to 80 in the convective envelope and from about 200 to 1000 when including the stable layer. The numerical resolution is $Nr=769$ in radius and $N_{\theta}=$ 512 or 1024 in latitude, with $N_{\phi}=2 N_{\theta}$ (a higher horizontal resolution for the low Rossby cases that develop smaller scale dynamics; see \citealt{Takehiro2020} for more details on the critical convection mode excitation). We now briefly summarise some of their key magnetohydrodynamical properties,
which were first analysed in \citep{Brun2022}.

\subsection{Rotation profiles and their link to magnetic properties: The role of the Rossby number}\label{sec:RossbyEvol}

In Fig.~\ref{fig:0} we display the evolutionary tracks of four stars that have stellar masses ranging from 0.5 to 1.1 $M_{\odot}$, starting from the PMS all the way to the TAMS; in other words, similar to the mass range used in the 3D MHD dynamo solutions used in this study. To do so, we plotted their evolution in a normalised Rossby number versus age diagram, using 1D stellar structure and evolution models computed with the Starevol code \citep{amard2020}. We superimposed the 15 ASH models on the plot, to show how our parameter space study can cover several temporal phases of evolution (see also \citealt{Emeriau-Viard2017} for a detailed specific study of the PMS phase). We used the following definition of the Rossby number:
\begin{equation}
 Ro_{\rm f} = |\omega|/2\Omega_*
 \label{eq:01}
\end{equation}
with $|\omega|$ being the mean vorticity of the convective flows, taken at the middle of the convection zone in the ASH simulation and stellar evolution tracks, $D$ the thickness of the convective envelope, and $\Omega_*$ the model rotation rate. This definition corresponds to the ‘fluid’ Rossby number, a measure of the influence of the Coriolis force on the non-linear advection term in the Navier-Stokes equation. The fluid and more usually stellar Rossby numbers can be related to one another, as is shown in Appendix~\ref{sec:appB} (see also \citealt{Brun2017a} for an overview of the different definitions). In Fig.~\ref{fig:0} we normalised it to the value of the Sun's $Ro_{\odot}$, here chosen to be 0.9. Before getting into the details of the figure, we wish to quickly recall how the Rossby number characterises the dynamics. For low values of the Rossby number, the rotational effects are dominant and force the dynamics to be aligned along the rotation axis (the so-called Taylor-Proudman constraint; \citealt{Pedlosky1987,Brun2002a,Miesch2006a,busse1983,Busse2006}). This usually results in an internal cyclindrical DR profile. For intermediate values, thermal effects via baroclinic torques can bend the iso-contours of $\Omega$ to be more conical at mid-latitudes \citep{Miesch2006a}, as in the Sun and its helioseismically inferred angular velocity with a fast equator and slow poles. For large values of the Rossby number, the rotational effects are weaker and the local angular momentum conservation can lead to anti-solar DR profiles, with a slow equator and fast poles \citep{Gastine2014b}. These different states of internal angular velocity profiles are represented by the symbols (cross, square, and circle) in Fig.~\ref{fig:0}.

We can notice several key pieces of information about our set of 15 ASH simulations that are plotted as symbols of various shapes and colours. In this stellar evolutionary diagram, we first see a clear trend of stars evolving from the bottom left towards the upper right. Indeed, as stars age they tend to slow down (these 1D stellar evolutionary tracks do not consider the possible stalling of the spin-down advocated by some authors, as is discussed in the introduction and Sects.~\ref{sec:RossbyEvol} and \ref{sec:3}). We note that the large range of Rossby numbers of the 3D MHD dynamo study covers a significant part of the stellar evolutionary track of solar-type stars. Some stars are in the low Rossby number regime (square symbols), whereas others are in the slow rotation regime (cross symbols). We also note that there is a continuous change in the associated DR regimes, going from banded or quenched in the early stages towards becoming anti-solar for a long-enough secular evolution. As of today, it is difficult to say if solar-type stars will become anti-solar before turning into sub-giant or red giant stars. In \citet{noraz2022a}, we conducted a systematic study of the ‘Kepler’ sample published in \citet{Santos2021} and found 22 possible candidates that would be worth observing further in order to put more constraints on slowly rotating stars. Nevertheless, the continuous transition of states is interesting in and of itself and could sustain a Sun in time scenario that describes a magneto-rotational dynamical evolution of stars like the Sun over secular time frames. Indeed, we also added in Fig.~\ref{fig:0}, the magnetic dynamo states of the simulations that were found in \citet{Brun2022}. For low Rossby numbers, most models harbour short cycle period dynamo actions. Periods of the order of half a year to two years are found in the models. For intermediate, more solar-like Rossby number values, corresponding to most of the main sequence of those stars, we find decadal cycle periods as in the Sun or 18Sco \citep{donascimentoHalelikeCycleSolar2023}. Finally, evolved old stars may lose their cyclic magnetic behaviour and display instead statistically stationary magnetic states, with very stable polarity in each hemisphere over secular time frames before turning into more evolved stages out of the main sequence, for which our 3D MHD study is not designed.

\begin{figure*}
    \sidecaption
        \includegraphics[width=12cm]{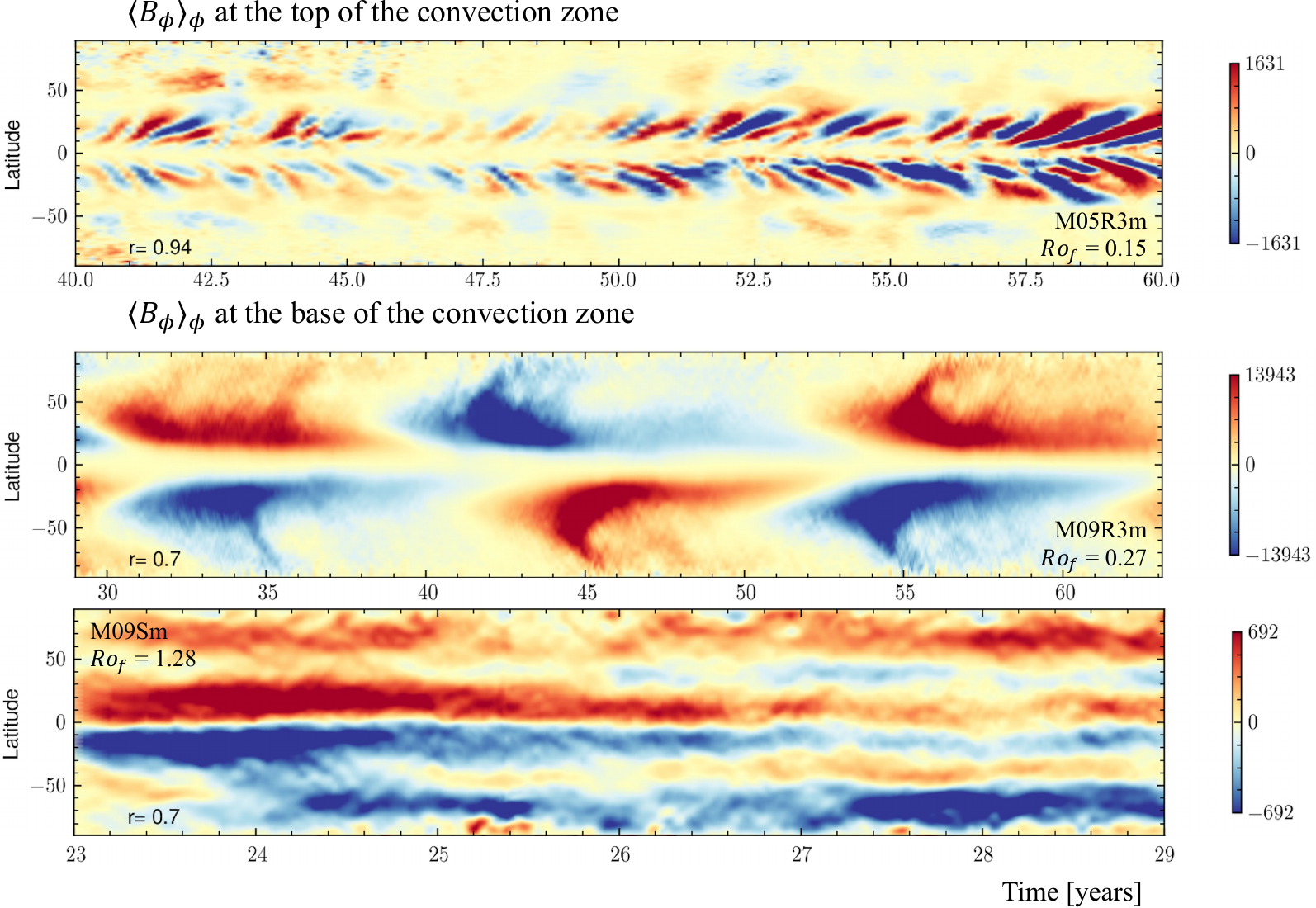}
         \caption{Typical magnetic butterfly diagrams (i.e. time-latitude diagram of the axisymmetric toroidal magnetic field) achieved in the simulations published in \citet{Brun2022}. On the top panel, the butterfly diagram is shown as a colour contour plot in Gauss for the rapidly rotating cases (low Rossby numbers), in the middle panel for intermediate value (Sun-like) and in the bottom panel for statistically steady solutions for slowly rotating stars with large Rossby numbers.}
         \label{fig:1}
\end{figure*}

\subsection{Magnetic butterfly diagrams versus stellar dynamo types}

In order to illustrate a bit more the dynamo states achieved in the solar-type stars modelled in the study of \citet{Brun2022}, we represent in Fig.~\ref{fig:1} three typical magnetic butterfly diagrams found in this 3D MHD parameter study. The butterfly diagrams were formed by azimuthally averaging the toroidal field of the simulation at any depth in the simulation (usually either in the tachocline at the base of the convection zone or near the surface) and by stacking these latitudinal bands in time to form time-latitude contour plots. Each of these diagrams covers several decades of evolution and is strikingly different at first sight.

In the top panel, we show a representative butterfly diagram near the surface for rapidly rotating stars, those with small Rossby numbers. We clearly see the small red-blue alternating colour bands at mid-latitudes, illustrating here local polarity inversions. The short cycle period is of the order of six months in the case illustrated. The bands propagate polewards, and 
the dynamo wave follows the Parker-Yoshimura rule for $\alpha-\Omega$ dynamo types (as was demonstrated in \citealt{Brun2022}). 

In the middle panel, we show the butterfly diagram near the base of the convection zone for the case with an intermediate Rossby number. We clearly see the long cyclic behaviour, with three consecutive cycles with a typical global polarity reversal, as is seen in the Sun; in other words, the magnetic features have reversed polarity in each hemisphere and the polarity swap signs from one cycle to the next, as \cite{haleMagneticPolaritySunSpots1919} reported from the Sun in his seminal paper. This is due to the dominance of the dipolar-antisymmetric dynamo mode, although some desynchronisation between the northern and southern hemispheres can be seen, which is due to a non-negligible quadrupolar-symmetric mode that leads to a more independent hemispherical magnetic response \citep{Gallet2009,Derosa2012}.
Another key feature of the middle panel butterfly diagram is the mid-latitude equatorward branch, and a high latitude polar branch. Unlike the fast rotating dynamo cases at a low Rossby number, these long period cycle dynamos do not follow the classical Parker-Yoshimura rule. Both $\alpha$ and $\Omega$ effects do play a role, but the dynamo loop leading to a cyclic behaviour involves the non-linear retroaction of the large-scale toroidal field on the large-scale shear \citep{Strugarek2017}. A new cycle starts with the reversed polarity, when the Maxwell stress modifies locally the sign of the gradient, $\partial \Omega/\partial \theta$. The longer cycle period comes from the time it takes for the field to alter the angular velocity shear, as this only occurs above a certain field strength. Indeed, the $\Omega$ effect is a linear field stretching mechanism that takes, in the specific case illustrated in Fig.~\ref{fig:1}, about ten years to act. We also find that this dynamo operates much deeper, straddling the base of the convective envelope, where significant energy transfers (up to several \% of the solar luminosity) allow global polarity reversals of the large-scale magnetic field in a prey-predator-type mechanism, generating torsional oscillations (see \citealt{Brun2022} for their analysis).

In the bottom panel, we also display the butterfly diagram near the base of the convection zone for a typical large Rossby number case ($Ro_{\rm f} > 1$). This type of dynamo possesses an anti-solar DR. As is shown in \citet{Noraz2021} (see also \citealt{Karak2020}), such reverse angular velocity profiles (with respect to the Sun) often yield stationary dynamos that do not show clear and systematic polarity reversals. Temporal variability is still present, with the large-scale toroidal magnetic wreaths exhibiting amplitude variations, sometimes not going all the way around the 360 degree longitudes of the star (see \citealt{Nelson2013}). Here, we do not find any sign of polarity reversal in the non-axisymmetric components of the field that were found by \cite{vivianiTransitionAxiNonaxisymmetric2018}.

Our simulations, despite their sophistication, may not faithfully reproduce every nuance of real stellar surface dynamics. Meridional circulation and torsional oscillations amplitudes observed in the Sun, for instance, may not be precisely mirrored in our simulations (see \citealt{hottaDynamicsLargeScaleSolar2023} for a review). Nevertheless, we do not aim to reproduce the precise solar case or every detail of solar surface magnetism in this parametric study, but instead wish to unveil broader trends as a function of the Rossby number over evolutionary timescales. We believe that the trends that will be discussed in Sects.~\ref{sec:3} and \ref{sec:4} are indicative of genuine energy and force balances occurring within solar-type star convective envelopes (as is discussed in \citealt{2013GeoJI.195...67D},  \citealt{aubertSphericalConvectiveDynamos2017a}, \citealt{Augustson2019}, and references therein).

\begin{figure*}
    \begin{center}
        \includegraphics[width=0.45\linewidth]{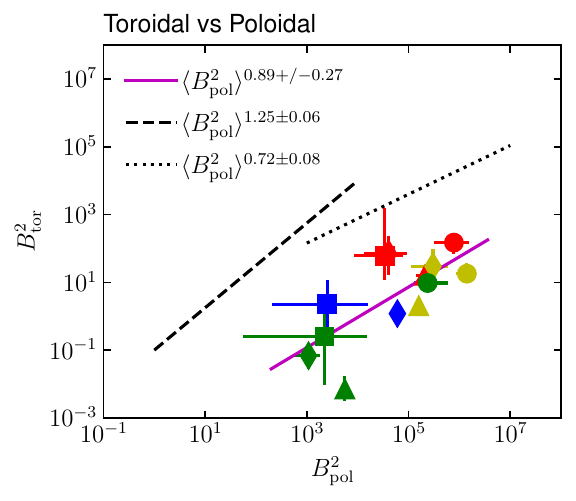}
        \includegraphics[width=0.45\linewidth]{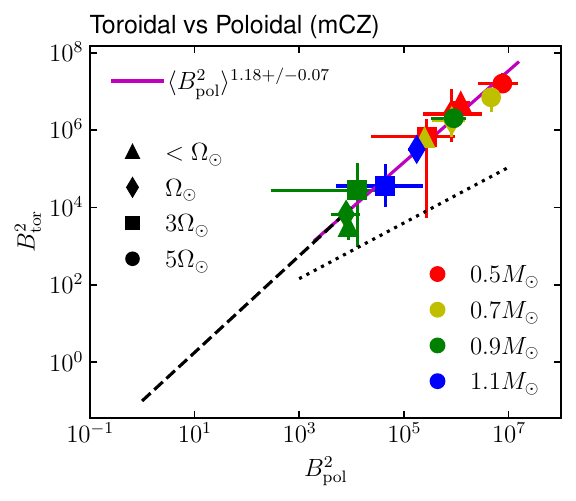}
    \end{center}
    \caption{{\it Left:} Magnetic ratio between the squared toroidal, $B_{tor}$, and poloidal, $B_{pol}$, components near the surface. Two observational fits (dashed black for $M>0.5M_\odot$ and dotted lines for $M<0.5M_\odot$) from \citet{See2015} are indicated, as well as the fit of the simulations (purple line) and their respective error bars. {\it Right:} Same ratio within the convective envelope. The coloured symbols have the following meanings: red, yellow, green, and blue represent stars with 0.5, 0.7, 0.9, and 1.1 $M_{\odot}$, respectively, and triangle, diamond, square, and filled circle shapes the rotation rates for, respectively, slow (the exact value depends on mass see \citealt{Brun2022}), one, three, and five times the solar rotation rate, $\Omega_{\odot}$. The error bars indicate the minimum and maximum values reached by the model during the integrated time (one or several cycle periods) used to compute the mean value indicated by the symbol.}\label{fig:2}
\end{figure*}

In summary, one can thus imagine that, as a solar-type star ages, it will respectively go through these three magnetic and rotation states. In order to further verify if such a stellar magneto-rotational scenario is plausible, we wish to compare other magnetic proxies with recent observations of stellar magnetism. To this end, we now turn to search in our dynamo database for various trends with respect to some global stellar parameters by splitting the field into various components (toroidal, poloidal, axisymmetric, dipolar, multipolar, etc...). In the following Sect., two simulations out of the set of 15 simulations presented above (namely M07R3m and M11R5m, see \citealt{Brun2022} for naming nomenclature) will not be considered because of a gap in the data needed at the time of the present study (spatial and temporal gaps, respectively). The top of the numerical domain is $r_{\rm top}=0.95$ R$_*$ for $M=5$ M$_\odot$ models and $r_{\rm top}=0.97$ R$_*$ for all the others. Values referred to as near the surface in the rest of the paper are evaluated at $r=0.9997$ $r_{\rm top}$ for $M=5$ M$_\odot$ models, $r=0.9993$ $r_{\rm top}$ for M11R3m, and $r=0.9998$ $r_{\rm top}$ for all the others.

\section{Magnetic dynamo trends with stellar parameters}\label{sec:3}

Having recalled the main broad properties of the set of dynamo solutions considered in the present paper (see \citealt{Brun2022} for more details), we now wish to look systematically at various trends regarding their magnetic properties with respect to key stellar parameters (such as the Rossby number, stellar mass,  or field geometry). In doing so, we intend to assess how well the set of stellar dynamo simulations can further confirm our Sun in time scenario, by directly comparing our results to those published in the observational studies of \citet{See2015,See2019,See2019b}. To that end, we use similar layouts for the figures to ease the direct comparison and plot observational scaling laws (fits) when available in the publications. We account for the difference in the Rossby number definition used in observational studies (stellar) and the present paper (fluid) when showing observational trends as a function of $Ro_{\rm f}\simeq Ro_{\rm s}/2.26$ (see Appendix~\ref{sec:appB} for more details).
\begin{figure}[!ht]
    \begin{center}
        \includegraphics[width=0.9\linewidth]{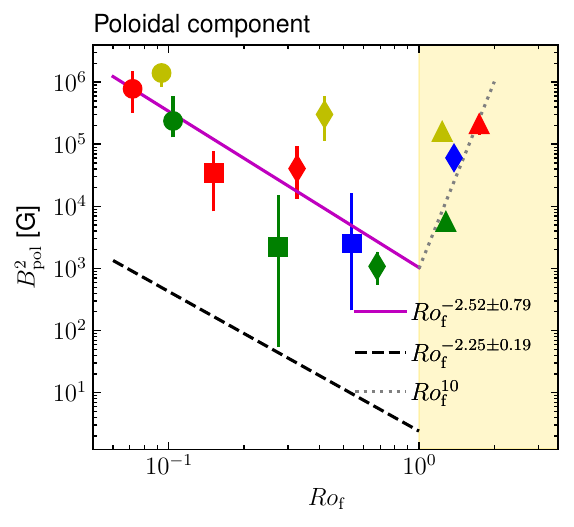}
    \end{center}
    \caption{Poloidal field component squared amplitude as a function of the Rossby number, $Ro_{\rm f}$. At low $Ro_{\rm f}$, the agreement between the simulations (purple line) and observation (dashed black line) is quantitatively good, with both showing a similar decreasing trend. At a higher Rossby number, $Ro_{\rm f}>1$ (yellow range), not covered by the observational database, an inverse trend is indicated by the simulations, and hence suggests a minimum in poloidal field strength near $Ro_{\rm f} \sim 1$, which could have interesting consequences for stellar spin-down via wind braking. An indicative trend proportional to $Ro_{\rm f}^{10}$ is shown using a dotted grey line. We note the V shape that the two trends (purple and dotted lines) form, with the minimum being near $Ro_f\sim 1 $. The coloured symbols have the same meanings as in Fig.~\ref{fig:2}.\label{fig:3}}
\end{figure}

\begin{figure*}
    \begin{center}
        \includegraphics[width=0.45\linewidth]{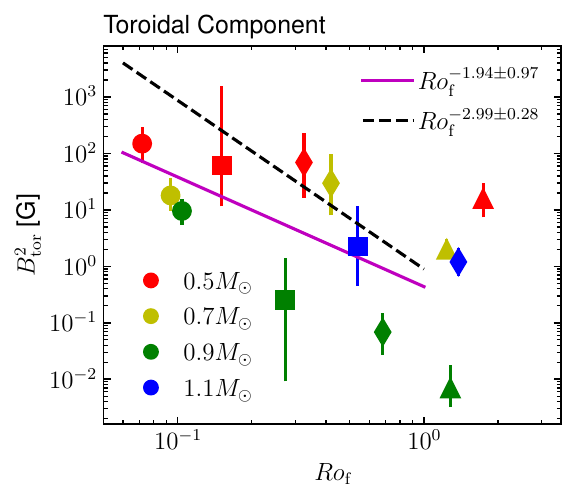}
        \includegraphics[width=0.45\linewidth]{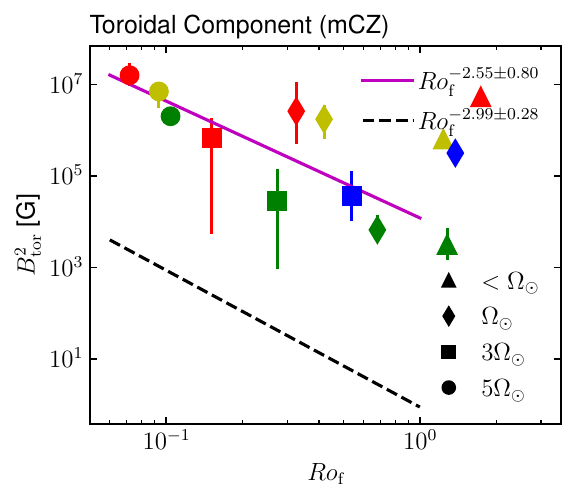}
    \end{center}
    \caption{{\it Left:} Toroidal field component squared amplitude as a function of the Rossby number, $Ro_{\rm f}$, probed near the surface. At low $Ro_{\rm f}$, the agreement between the simulations (purple line) and observational (dashed black line) fits is qualitatively good, with both simulations and observations showing a decreasing trend, although the field amplitude is too low near the surface (left panel). This is likely due to our choice of top magnetic boundary conditions. {\it Right:} Same quantity probed in the middle of the convection zone, where the agreement is quantitatively good, with both fits being close in terms of the power law index. A possible inverse trend is indicated by the simulations at higher Rossby numbers, $Ro_{\rm f}>1$ (not covered by the observational database), and hence suggests a minimum in the toroidal field strength near $Ro_{\rm f} \sim 1$, which could have interesting consequences for stellar spin-down via wind braking.\label{fig:4}}
\end{figure*}

\subsection{Poloidal versus toroidal magnetic field properties}

In Fig.~\ref{fig:2} we display the magnetic ratio, defined here as the relation between the toroidal and the poloidal field amplitude squared. We show this relation near the top of the dynamo solution (left panel) and at the middle of the convective envelope (right panel) for each stellar spectral type.
The definitions of the poloidal, $B_{pol}$, and toroidal magnetic field, $B_{tor}$, can be found in Appendix~\ref{sec:appA}. To ease direct comparisons to observations, we truncated the spherical harmonics decomposition, conserving only $\ell\leq 5$ in the computation of both quantities (see Appendix~\ref{sec:appA}, and also \citealt{Vidotto2016, See2019}).

Looking first at the result near the surface, we see that the amplitude of the toroidal field is smaller in the set of simulations.
We note that the simulation trend, $B_{tor}^2 \propto (B_{pol}^2)^n$ (purple fit), lies in between the two values proposed by \citet{See2015}, and hence captures the mean trend found in the observational studies. The lower amplitude may be explained by the potential field surface magnetic boundary condition of the simulation, enforcing a zero longitudinally averaged magnetic toroidal field. Even though we probed that interdependence, a few mesh points below the top of the numerical domain (so not quite where the zero value is enforced numerically), this is likely to still have an influence on the ratio between the two field geometries. In order to quantify this effect, we now show in the right panel of Fig.~\ref{fig:2} this relation in the middle of the convection zone, where this influence from the boundary condition does not hold any more. We indeed note that the simulations better match the data in terms of the amplitude reaching a value of $B_{tor}^2$ around $10^4 G^2$ and above. The simulation trend matches the higher value of the exponent, $n$ (i.e 1.18 vs 1.25); in other words, the simulations tend to qualitatively agree with the observations, which is pretty encouraging. If we further believe the scenario that starspots are created by the surface emergence of flux ropes coming from deeper in the convection zone (as a whole entity or subparts of it), then the observed relation may well represent the toroidal geometries of the deeper interior, as is seen here (see also \citealt{finleyHowWellDoes2024} for further investigations into the M11R3m model of this set). Since in the simulation we do not have yet the formation of large compact magnetic features due to the lack of local resolution or near-surface dynamics, and the choice of potential-field magnetic boundary conditions, the near-surface ratio in the left panel may be biased in the simulation to have much lower near-surface toroidal fields. Hence, assessing this ratio in the middle of the convection zone of the simulations is less affected by the potential field boundary conditions than at the surface, which is confirmed by comparing the two panels of Fig.~\ref{fig:4}. Work is in progress to add a realistic atmosphere on top of current global dynamos to have much improved surface magnetic field boundary conditions \citep{Warnecke2016,Perri2021,delorme2022,2022MNRAS.517.2775K}.

\subsection{Trends with the Rossby number}

We now turn to considering various trends of the magnetic field and its components with the Rossby number, $Ro_{\rm f}$.

\subsubsection{Poloidal and toroidal decomposition}

In Fig.~\ref{fig:3} we show how the near-surface poloidal magnetic field squared amplitude depends on the Rossby number. At low Rossby numbers ($Ro_{\rm f} <1$) where the observational data are concentrated, the agreement with the observations is quantitatively good regarding the tendency. Another interesting property can be seen in Fig.~\ref{fig:3} for large Rossby number values. We see that the trend is opposite in sign, with $B_{pol}^2$ now increasing with $Ro_{\rm f}$ rather than decreasing as for the more rapidly rotating (low $Ro_{\rm f}$) dynamo cases. This is due to a sharp transition in the DR in the model, going from solar to anti-solar dynamo (see \citealt{Matt2011,Gastine2014b,Brun2017,Brun2022}). This is a very interesting property that needs to be studied further, as this possible V-shape trend could explain a weaker temporary wind braking, due to a minimum in the poloidal field strength as a function of rotation (Rossby number). \citet{Brandenburg2018} and \citet{lehmannSPIRouRevealsUnusually2023} seem also to find a reverse trend for the magnetic flux amplitude of slowly rotating stars in their observational study. Clearly, studying the high Rossby number states is becoming very timely, and observational investigations already started \citep{noraz2022a,donatiMagneticFieldsRotation2023,cristofariMeasuringSmallscaleMagnetic2023}.

\begin{figure*}[!th]
    \begin{center}
        \includegraphics[width=0.31\linewidth]{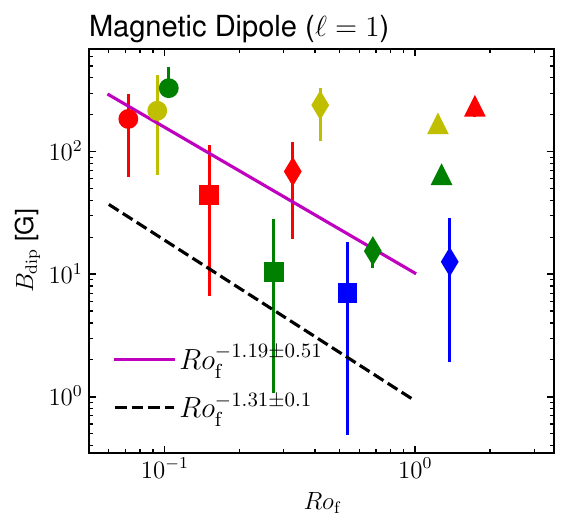}
        \includegraphics[width=0.31\linewidth]{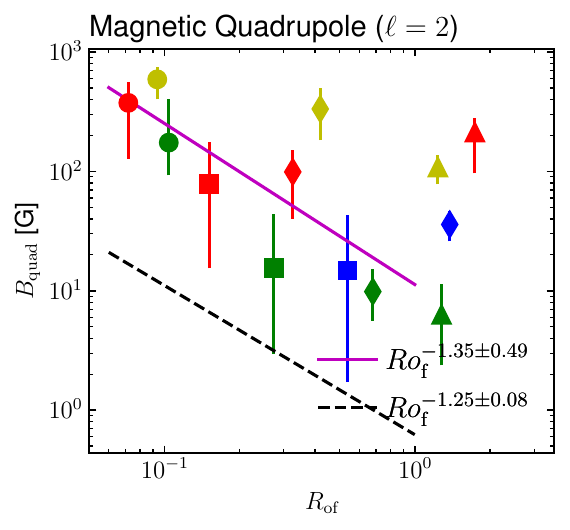}
        \includegraphics[width=0.31\linewidth]{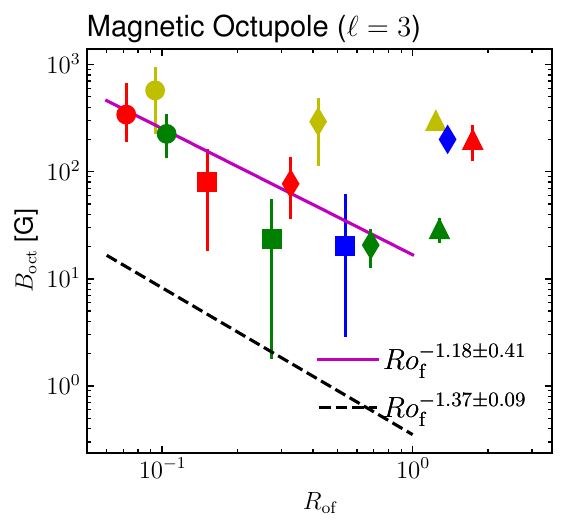}
    \end{center}
    \caption{Multipolar decomposition of the magnetic field near the surface (with $m$ components summed quadratically), showing the dipole, $\ell=1$ ({\it Left}), quadrupole, $\ell=2$ ({\it Middle}), and octupole, $\ell=3$ ({\it Right}), as a function of the Rossby number, $Ro_{\rm f}$. At low $Ro_{\rm f}$, the agreement between the simulations (purple line) and observational (dashed black line) fits is quantitatively good for all low-degree dynamo modes, with all modes showing a decreasing trend and power laws comparable to observations.} The possible inverse trend for high Rossby numbers is also found in all three multipoles. Symbols have the same meanings as in Fig.~\ref{fig:4}.\label{fig:5}
\end{figure*}

Turning now to the toroidal component, we show in Fig.~\ref{fig:4} the dependency of $B_{tor}^2$ on $Ro_{\rm f}$ near the surface (left panel) and deeper inside the convection zone (right panel). Near the surface, as for Fig.~\ref{fig:1} (left panel), the simulations only broadly match the observed properties. The field values are a bit too low and have only in the large the same declining trend for low $Ro_{\rm f}$. The situation improves significantly when forming the same figure deeper down in the simulation and focusing on the low $Ro_{\rm f}$ part of the plot. We see that the simulation trend (purple fit) agrees better with the observations. The field amplitude is a bit too high, indicating that the toroidal field at the stellar surface is likely weaker than deep in the stellar dynamo, but not as weak as the potential field boundary condition imposes (see left panel). We also notice the inverse trend for high $Ro_{\rm f}$, which we will need to look into in the near future \citep{noraz2022a}. Very little observations are yet reporting toroidal field measurements for high Rossby number stars, except for a few M dwarfs in \cite{donatiMagneticFieldsRotation2023} and \cite{lehmannSPIRouRevealsUnusually2023}, not quite directly comparable with our study, which is more focused on G and K-type dwarfs.

Indeed, the errors from the simulations fits are not as small as in the observations due to the moderate number of dynamo models compared to the number of observed stars, but overall the fits agree. It should be noted that running these 15 3D MHD global convective dynamo simulations over many decades of physical time is already a challenge that took several years on massively parallel supercomputers, and that nothing in particular was done or tuned in the simulations to get this comprehensive match with the observations. This is reassuring and reinforces our confidence in the set of simulations published in \citet{Brun2022} and further discussed here.

\subsubsection{Multipolar decomposition}

It is interesting to further study the properties of the magnetic field of the dynamo simulations by considering the behaviour of single low $\ell$ spherical harmonics degree magnetic field components such as the dipole, quadrupole, and octupole ($\ell = 1, 2, 3$), which can be observed in most ZDI studies of magnetic stars \citep{Petit2008,Marsden2014,See2015,See2019}. We do so in Fig.~\ref{fig:5}, using the definition listed in Appendix~\ref{sec:appA}. 

The dipole is the dominant magnetic ingredient for efficient wind braking (then the quadrupole; \citealt{Reville2015a,Finley2017}), and thus being able to predict its amplitude is key when trying to understand the magneto-rotational evolution of solar-type stars. Looking at the leftmost panel of Fig.~\ref{fig:5}, we report a good quantitative agreement for the low Rossby number values in terms of trend and amplitude. Again, for large values we see that the dipolar field increases for slowly rotating stars. This confirms that such a V-shaped dip in the trend could play as a minimum in wind braking efficiency at this Rossby number transition, explaining a possible stalling or weakening of solar-type star spin-downs at an intermediate age \citep{VanSaders2016,Curtis2020}.

Turning to the middle and right panels of Fig.~\ref{fig:5}, we see that both the quadrupolar and octupolar magnetic field components are also in good quantitative agreement with the observational trends. Both purple fit indexes match the observational trends of \citet{See2019b} at low Rossby numbers. They show an inverse trend for large $Ro_{\rm f}$ values too, reinforcing the dipolar trend discussed above and explaining why it is also clearly seen in the poloidal component in Fig.~\ref{fig:3}.

\begin{figure*}
    \begin{center}
        \includegraphics[width=0.30\linewidth]{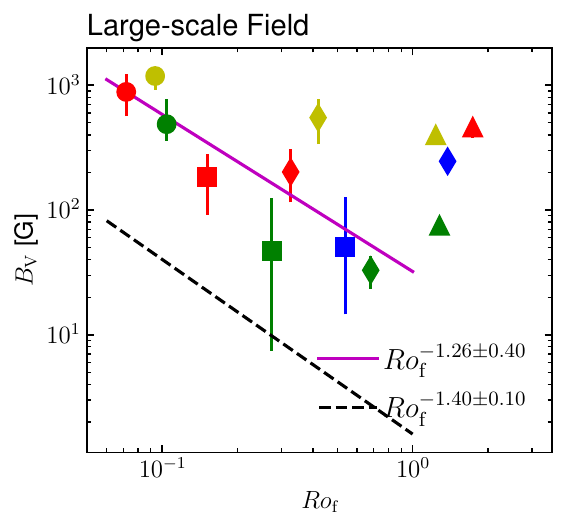}
        \includegraphics[width=0.33\linewidth]{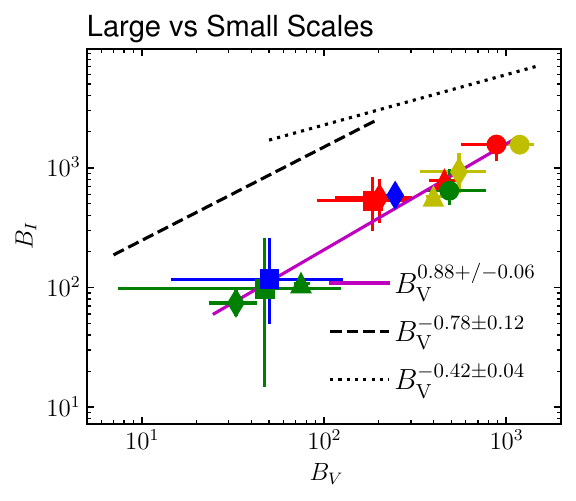}
        \includegraphics[width=0.33\linewidth]{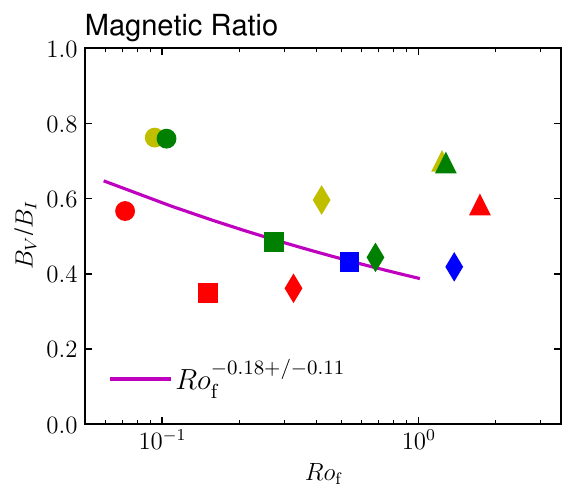}
    \end{center}
    \caption{Various trends for our theoretical proxies of both the Stokes V magnetic field measurement, $B_V$, and the total field measurement via Zeeman spectroscopy, $B_I$. {\it Left:} The $B_V$ dependency on the Rossby number is well captured, with both fits (purple for the simulations and dashed black lines for the observations) agreeing quantitatively well in terms of the power law index. {\it Middle:} Magnetic relation between $B_V$ and $B_I$ showing, as in Fig.~\ref{fig:2}, an expected positive trend. {\it Right:} $B_V/B_I$ ratio as a function of $Ro_{\rm f}$, with a fit (purple line) indicating a weak dependency. Symbols have the same meanings as in Fig.~\ref{fig:7}.} \label{fig:6}
\end{figure*}

\begin{figure}
    \begin{center}
        \includegraphics[width=0.9\linewidth]{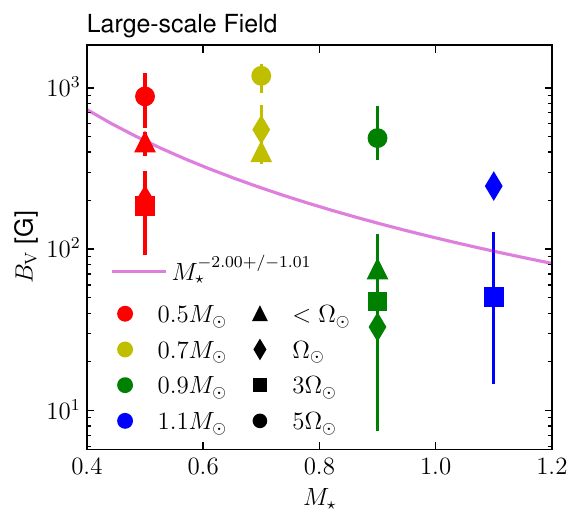}
    \end{center}
    \caption{Theoretical proxy for Stokes V magnetic field, $B_V$, vs stellar mass in the simulation's dataset. Symbols have the same meanings as in Fig.~\ref{fig:2}. We note a rather strong decreasing trend with stellar mass (purple fit exponent around -2), the rotation rate leading to some spread, and the relatively large error bar on the fit. }\label{fig:7}
\end{figure}

\subsubsection{Large-scale versus total magnetic field relationships}

We now wish to consider how the total magnetic field, $B_I$, and the large-scale field, $B_V$, obtained by different techniques based on the Zeeman effect, behave with respect to one another.
We already showed in \citet{Brun2022} that filtering the surface magnetic field of the simulations was necessary for a meaningful direct comparison with observations. Indeed, standard equipartition dynamo scaling \citep{davidson2014,Augustson2019}, considering the bulk magnetic field, differs from the one derived using the large-scale surface magnetic field.
Comparing the observed $B_V$ obtained with ZDI techniques with respect to $B_I$ obtained by Zeeman spectroscopy can help us disentangle the contribution of the large-scale and smaller-scale magnetic fields (down to observational limit of distant stars) to the overall dynamo mechanism occurring inside solar-type stars.
We defined our proxy Stokes V magnetic field, $B_V$, as the filtered low $\ell$ degree magnetic field (up to $\ell_{max}= 5$, see Appendix~\ref{sec:appA} and \citealt{See2019}). We defined our proxy Zeeman total magnetic field, $B_I$, as the normalised near-surface integrated magnetic field, keeping all degrees $\ell$ (see Appendix~\ref{sec:appA}). The distinction between these two magnetic field definitions helps to characterise the relative sensitivities of the large-scale field and the total field to stellar parameter changes. 
In Fig.~\ref{fig:6} we present the dependency of $B_V$ on $Ro_{\rm f}$ (left panel), $B_V$ vs $B_I$ (middle panel), and $B_V$/$B_I$ vs $Ro_{\rm f}$ in the rightmost panel.

In the left panel of Fig.~\ref{fig:6} we also report a good quantitative agreement between the simulation and observational fits. We find a decreasing trend for the low Rossby number region of the plot, which matches the observational power law index of \citet{See2019} with $Ro_{\rm f}$. We also note the reverse trend for high $Ro_{\rm f}$, similar to the ones shown for low-degree components in Fig.~\ref{fig:5}, as could be anticipated, given their close relationship. In the middle panel, we plot the total field $B_I$ as a function of $B_V$, in a similar way to what we did in Fig.~\ref{fig:2} for the poloidal and toroidal components. The trend is clear; both fields are positively correlated and not quite linearly related. We also note that $B_V$ is larger and $B_I$ smaller than the typical amplitude expected from observations, both near the surface and deeper in the convection zone (the latter is not shown here). This is likely to be a limitation of the large eddy simulation (LES) approach adopted in these simulations, whereby the dynamics of the smallest scales is not resolved. Nevertheless, the trends we find are robust, implying that the magnetic field energy distribution (spectrum) between large- and small-scale fields is expected to evolve as the star ages. In that context, we finally show in the rightmost panel how the ratio $B_V/B_I$ behaves with respect to the Rossby number. This is to verify if, as the star rotate slower and slower, there is a tendency to build less large-scale magnetic fields. There is a weak trend, stating that indeed the large fields tend to diminish slightly faster than the total field, implying a busier and smaller-scale magnetic field near stellar surfaces. Such a broad tendency can be seen in the observational data of \cite{See2019} (see their Figure 2) for $M>0.5M_\odot$; however, the number of stars considered is also small, and any stronger conclusion will need further investigations on both theoretical and observational sides. Again, when $Ro_{\rm f}$ goes over 1 and the DR of the star flips direction, now harbouring a slow equator and fast poles, the situation reverses and stronger large-scale magnetic fields are generated by the dynamo. Hence, there is still a peculiar region both in the field geometry and the amplitude near $Ro_{\rm f} \sim 1$, which we will investigate in the near future both observationally and theoretically. 

\begin{figure*}
    \begin{center}
        \includegraphics[width=0.45\linewidth]{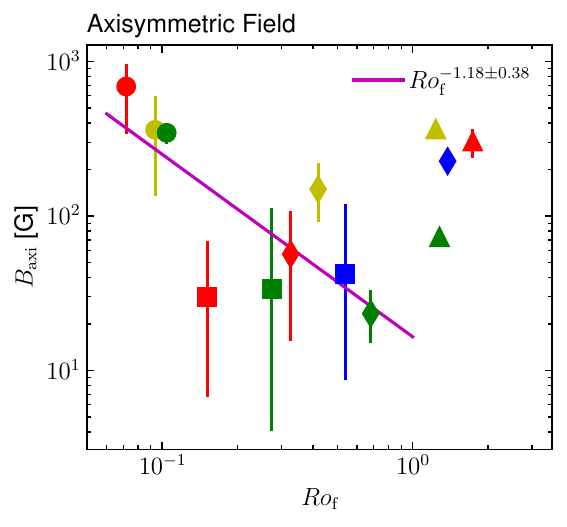}
        \includegraphics[width=0.45\linewidth]{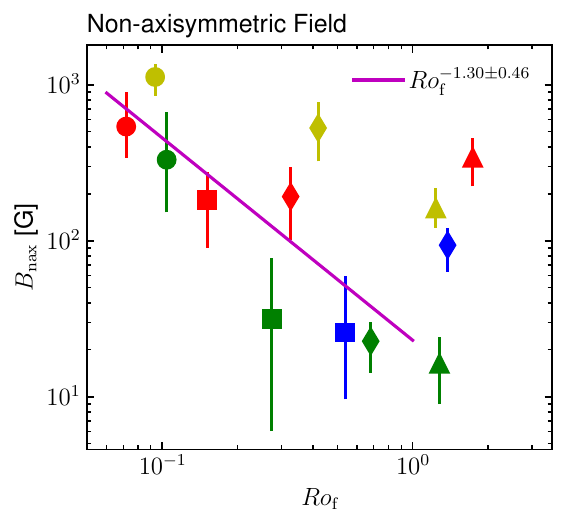}
    \end{center}
    \caption{Axisymmetric ($m=0$, {\it Left}) and non-axisymmetric ($m\neq 0$, {\it Right}) magnetic field decomposition variations with the Rossby number. No clear differences are seen between the two field components; the purple fit exponents in the low Rossby number regions being close and negative, with at best a tendency for the field to be more axisymmetric for larger Rossby number values. An inverse trend is also seen for large Rossby numbers, as was expected from the previous analysis. Symbols have the same meanings as in Fig.~\ref{fig:7}.}\label{fig:8}
\end{figure*}

In Fig.~\ref{fig:7}, we display how the large-scale magnetic field, $B_V$, scales as a function of the stellar mass. In the study, we cover four mass bins that can help us search for possible trends. We see that indeed there is a decreasing field amplitude with stellar mass, with the fit indicating a high negative exponent, -2, but with a relatively large error bar due to some rotational spreading. This result could seem counterintuitive, as more massive stars are more luminous (recall that for solar-type stars $L_* \propto M_*^4$; \citealt{hansen1995}). However, this effect is compensated for by the fast narrowing of the convective envelope with stellar mass, resulting in a much lower averaged density in the convective envelope, and hence a lower kinetic energy reservoir \citep{Brun2017a,Brun2022}. Observations by \citet{johns-krull2000} also found that the equilibrium field, $B_{eq}$, in M dwarfs possesses a higher amplitude when compared to that of G or F stars. They used the definition of the equilibrium field as satisfying $P_{mag}\sim B_{eq}^2=P_{gaz}\sim \bar{\rho}T_{eff}$ at the surface \citep{Brun2015c} and they found a relatively good agreement with their measurements for the same reason -- that $T_{eff}$ varies less than the mean density, $\bar{\rho}$, when going from stellar spectral M to F.

\subsubsection{Axi- versus non-axisymmetric magnetic field trends}

In Fig.~\ref{fig:8} we show how the axisymmetric ($m=0$; left panel) and non-axisymmetric ($m\neq=0$; right panel) magnetic field components behave with respect to the Rossby number. We wish to see if the field tends to be less regular and axisymmetric under some conditions.

In the low Rossby number range, we note that the field geometry tends to keep its axi- versus non-axisymmetric nature. There is a slightly larger exponent of the simulation fit for the non-axisymmetric magnetic field component, although much attention has to be paid to the relevance of such a power law fit in the case of the axisymmetric field near the surface for low $Ro_{\rm f}$. This tends to indicate that the rotation near $Ro_{\rm f} \sim 1$ favours relatively more regular axisymmetric magnetic fields in the simulations. This trend is more pronounced for larger values of $Ro_{\rm f}$. We clearly see that axisymmetric field is particularly enhanced for the faster rotating cases ($5\Omega_\odot$, circle shapes). We also note that the triangle symbols, which again present the reverse trends, are higher in amplitude in the axisymmetric part, indicating a more symmetric magnetic field. This can easily be understood by returning to the bottom panel of Fig.~\ref{fig:1}, where we display a butterfly diagram for a typical slowly rotating case. We see that these types of models, possessing an anti-solar DR, usually develop large-scale, statistically steady magnetic wreaths in both hemispheres. Such wreaths are intricate and intertwined large-scale magnetic field ribbons that can be stable over relatively long periods of time \citep{Brown2010a}. They are dominantly axisymmetric, even though they are known to have their own complex dynamics as the degree of turbulence of the simulation is increased \citep{brownRapidlyRotatingSuns2008,Nelson2013}.

\section{Discussion and conclusion}\label{sec:4}

We have discussed how well a set of 15 3D MHD global dynamo solutions compare to stellar magnetism observations of solar-type stars. Overall, we find a very good quantitative agreement between our set of 15 simulations and the observational studies from the {\rm Bcool consortium} \citep{Marsden2014} published in \citet{See2015,See2019,See2019b}. In this ensemble study, several interesting features have been found.
For instance, we find that for low Rossby number values, $Ro_{\rm f} < 1$, the various trends in the range $Ro_{\rm f} = [0.05,1]$ imply a decreasing magnetic field amplitude with a relatively steep slope. The poloidal, toroidal, or multipolar decompositions all follow relatively clear decreasing trends with $Ro_{\rm f}$, confirming that young, fast-rotating stars tend to have a larger field amplitude than their older counterparts.
We also note that the near-surface toroidal magnetic field amplitude is too low in the simulations. We believe this is due to our choice of potential field magnetic boundary condition, which sets the toroidal field to zero at the top of the numerical domain. When comparing the toroidal magnetic field strength deeper down in the simulations, 
we recover a better agreement with the observations. This somewhat confirms that there is a clear link between the dynamo mechanism operating deep in the convective envelope 
and the subsequent emergence of magnetic fields on the surfaces of the observed solar-type stars.

An interesting result to keep in mind is a possible reversal of the magnetic trend when reaching higher Rossby numbers, $Ro_{\rm f}>1$, pointing towards an enhancement of the stellar magnetism as the rotational influence decreases. This change of trend suggests a minimum of the large-scale field amplitude around $Ro_{\rm f}\sim 1$, forming a V shape, which could have interesting consequences for stellar spin-down via wind braking. If such a behaviour is indeed occurring in stellar dynamos, then stars reaching this rotational state could potentially be in a minimum state of wind braking (due to the local weakening of the large-scale magnetic field (dipole) in the parameter space). This in turn could explain some observational claims advocating for a stalling scenario of stellar spin-down \citep{VanSaders2016,Curtis2020,Hall2021}, deviating from classical gyrochronology rotational laws \citep{Skumanich1972,Barnes2003}. Some hints that a reverse trend of the magnetic flux for slow rotators may exist have been put forward in \citet{Brandenburg2018}. While the existence of a minimum in the field amplitude near $Ro_f \sim 1$ is clear in our study, this does not lead to a full stop of the stellar spin-down process, but rather to a significant slowing down of the spin-down process that could then be revived once the star has slowly but surely crossed the $Ro_f>1$ transition. Such a deceleration of the rotational evolution could be further amplified if a decrease in the mass-loss rate is also happening \citep{metcalfe2022}. However, observational constraints on the high Rossby range, $Ro_{\rm f}>1$, are still limited as the sensitivity of current observational techniques decreases drastically for slow rotations \citep{Donati2006,benomarAsteroseismicDetectionLatitudinal2018}. The possible disappearance of starspots due to a change in the dynamo nature would further make observational characterisations of high Rossby targets difficult with photometric techniques. In that context, more study of stellar magnetism in stars near the $Ro_{\rm f}=1$ transition must be undertaken to confirm that a transition exists. In \citet{noraz2022a}, we propose a list of slowly rotating (probably anti-solar) stellar candidates, and we hope to be able to study their magnetic properties in the near future and confront it with our stellar magneto-rotational scenario. In that respect, the first large-scale magnetic field quantifications for high Rossby M-dwarfs were recently reported \citep{donatiMagneticFieldsRotation2023,lehmannSPIRouRevealsUnusually2023}, and we expect to see them soon for the solar-like G-K stars modelled in the present paper.

While very encouraging, this study could be improved in several ways. For instance, the Reynolds number of the dynamo simulations presented in this study have low to intermediate values (see \citealt{Brun2022}, Table 2), and as such are only numerical experiments to try to understand the complex and highly non-linear nature of stellar dynamos. Each of the individual simulations could possibly be an approximate realisation of the real single star it is supposed to represent. 
Indeed, global simulations of rotating convection still struggle to perfectly reproduce the solar case. In particular, there is currently a mismatch between global convection simulations and helioseismic observations regarding the power contained in giant convection cells, known as the ‘convective conundrum’ \citep{omaraVelocityAmplitudesGlobal2016,hottaDynamicsLargeScaleSolar2023}. This results in a slightly too large effective Rossby number in global convection simulations of the solar rotation rate. In this context, relative comparisons between models with different Rossby numbers can be done, but the absolute positioning of a given solar-type star should be considered with care. However, the comprehensive agreement found between \cite{Brun2022} and \cite{Strugarek2017}, using intrinsically different numerical methods, makes us confident about the relevance of these dynamo solutions in exploring and discussing the physical nature of solar-like cyclic activity, and the robustness of the overall trends found in our study. It is now important to reiterate that our primary focus is identifying and elucidating overarching trends. We intentionally avoided fine-tuning our simulations so as to make them agree with every observational detail. We thus find it particularly encouraging that such overall trend agreements arise without ad hoc adjustments. The existence in the simulation data of a magnetic upsurge in the high Rossby number regime then becomes a promising avenue for future research. For instance, the possibility of a significant change in the dynamo process in this high Rossby number regime will need further investigation \citep{noraz2022a,donatiMagneticFieldsRotation2023,cristofariMeasuringSmallscaleMagnetic2023,lehmannSPIRouRevealsUnusually2023}.

This relatively successful ab initio approach is encouraging for future studies that will introduce higher degrees of turbulence (Reynolds numbers) and more realistic surface boundary conditions \citep{Perri2021,delorme2022,2022MNRAS.517.2775K}. To summarise, this study suggests that a coherent magneto-rotational scenario for solar-type stars over secular evolution time, as is summarised in Fig.~\ref{fig:0} and in Sects.~\ref{sec:RossbyEvol} and \ref{sec:3}, is plausible. That is, the young stars start rotate relatively fast and are very active with short period magnetic cycles. Then, as they age and reach intermediate Rossby number values, their cycle period increases, reaching decade-long time spans due to a change in the dynamo operation. Near the $Ro_{\rm f} \sim 1$ transition, the stars may undergo a weakening of their wind braking, with what could appear to be a stall. As the stars age and continue to spin down more slowly, they may reach high Rossby number values and reverse their angular velocity profile. This new state of internal rotation modifies the stellar dynamo once more, leading possibly to a loss of its cyclic behaviour and the building of stronger large-scale magnetic fields, resulting in a revival of stellar spin-down.

\begin{acknowledgements}
All authors are thankful to L. Amard for providing the 1D stellar evolution models used to make Fig 1, P. Petit for useful discussions regarding the {\rm Bcool consortium} observations of stellar magnetism, V. See, T.S. Metcalfe, and the anonymous referee for his/her review. We acknowledge GENCI project 1623 for access to supercomputing infrastructures where most of these 15 simulations were computed. We acknowledge funding support by ERC Whole Sun \#810218, INSU/PNST grant and Solar Orbiter and Plato CNES financial supports. This work benefited from discussions within international teams “The Solar and Stellar Wind Connection: Heating processes and angular momentum loss” and "Solar and Stellar Dynamos: a New Era" supported by the International Space Science Institute (ISSI). A.S.B and A.S. are thankful to S.H. Saar for fruitful discussions.
\end{acknowledgements}

\bibliographystyle{aa}
\bibliography{SunInTime}

\begin{thebibliography}{83}
\expandafter\ifx\csname natexlab\endcsname\relax\def\natexlab#1{#1}\fi

\bibitem[{Ahuir {et~al.}(2020)Ahuir, Brun, \& Strugarek}]{Ahuir2020}
Ahuir, J., Brun, A.~S., \& Strugarek, A. 2020, Astronomy \& Astrophysics, 170, 1

\bibitem[{Amard \& Matt(2020)}]{amard2020}
Amard, L. \& Matt, S.~P. 2020, ApJ, 889, 108

\bibitem[{Aubert {et~al.}(2017)Aubert, Gastine, \& Fournier}]{aubertSphericalConvectiveDynamos2017a}
Aubert, J., Gastine, T., \& Fournier, A. 2017, Journal of Fluid Mechanics, 813, 558

\bibitem[{Augustson {et~al.}(2019)Augustson, Brun, \& Toomre}]{Augustson2019}
Augustson, K.~C., Brun, A.~S., \& Toomre, J. 2019, The Astrophysical Journal, 876, 83

\bibitem[{Ayres(1997)}]{ayres1997}
Ayres, T.~R. 1997, J. Geophys. Res., 102, 1641

\bibitem[{Barnes(2003)}]{Barnes2003}
Barnes, S.~A. 2003, The Astrophysical Journal, 586, 464

\bibitem[{Benomar {et~al.}(2018)Benomar, Bazot, Nielsen, Gizon, Sekii, Takata, Hotta, Hanasoge, Sreenivasan, \& {Christensen-Dalsgaard}}]{benomarAsteroseismicDetectionLatitudinal2018}
Benomar, O., Bazot, M., Nielsen, M.~B., {et~al.} 2018, Science, 361, 1231

\bibitem[{Brandenburg \& Giampapa(2018)}]{Brandenburg2018}
Brandenburg, A. \& Giampapa, M.~S. 2018, The Astrophysical Journal Letters, 855, L22

\bibitem[{Brown {et~al.}(2008)Brown, Browning, Brun, Miesch, \& Toomre}]{brownRapidlyRotatingSuns2008}
Brown, B.~P., Browning, M.~K., Brun, A.~S., Miesch, M.~S., \& Toomre, J. 2008, The Astrophysical Journal, 689, 1354

\bibitem[{Brown {et~al.}(2010)Brown, Browning, Brun, Miesch, \& Toomre}]{Brown2010a}
Brown, B.~P., Browning, M.~K., Brun, A.~S., Miesch, M.~S., \& Toomre, J. 2010, Astrophysical Journal, 711, 424

\bibitem[{Brun \& Browning(2017)}]{Brun2017}
Brun, A.~S. \& Browning, M.~K. 2017, Living Reviews in Solar Physics, 14, 4

\bibitem[{Brun {et~al.}(2015)Brun, Browning, Dikpati, Hotta, \& Strugarek}]{Brun2015c}
Brun, A.~S., Browning, M.~K., Dikpati, M., Hotta, H., \& Strugarek, A. 2015, Space Science Reviews, 196, 101

\bibitem[{Brun {et~al.}(2004)Brun, Miesch, \& Toomre}]{Brun2004}
Brun, A.~S., Miesch, M.~S., \& Toomre, J. 2004, The Astrophysical Journal, 614, 1073

\bibitem[{Brun {et~al.}(2022)Brun, Strugarek, Noraz, Perri, Varela, Augustson, Charbonneau, \& Toomre}]{Brun2022}
Brun, A.~S., Strugarek, A., Noraz, Q., {et~al.} 2022, The Astrophysical Journal, 926, 21

\bibitem[{Brun {et~al.}(2017)Brun, Strugarek, Varela, Matt, Augustson, Emeriau, DoCao, Brown, \& Toomre}]{Brun2017a}
Brun, A.~S., Strugarek, A., Varela, J., {et~al.} 2017, The Astrophysical Journal, 836, 192

\bibitem[{Brun \& Toomre(2002)}]{Brun2002a}
Brun, A.~S. \& Toomre, J. 2002, The Astrophysical Journal, 570, 865

\bibitem[{Busse(1983)}]{busse1983}
Busse, F.~H. 1983, Geophysical and Astrophysical Fluid Dynamics, 23, 153

\bibitem[{Busse \& Simitev(2006)}]{Busse2006}
Busse, F.~H. \& Simitev, R.~D. 2006, Geophysical and Astrophysical Fluid Dynamics, 100, 341

\bibitem[{Clune {et~al.}(1999)Clune, Elliott, Miesch, Toomre, \& Glatzmaier}]{Clune1999}
Clune, T.~C., Elliott, J.~R., Miesch, M.~S., Toomre, J., \& Glatzmaier, G.~A. 1999, Parallel Computing, 25, 361

\bibitem[{Cranmer \& Saar(2011)}]{cranmerTESTINGPREDICTIVETHEORETICAL2011}
Cranmer, S.~R. \& Saar, S.~H. 2011, The Astrophysical Journal, 741, 54

\bibitem[{Cristofari {et~al.}(2023)Cristofari, Donati, Moutou, Lehmann, Charpentier, Fouqu{\'e}, Folsom, Masseron, Carmona, Delfosse, Petit, Artigau, Cook, \& {the SLS consortium}}]{cristofariMeasuringSmallscaleMagnetic2023}
Cristofari, P.~I., Donati, J.-F., Moutou, C., {et~al.} 2023, Monthly Notices of the Royal Astronomical Society, 526, 5648

\bibitem[{Curtis {et~al.}(2020)Curtis, Ag{\"u}eros, Matt, Covey, Douglas, Angus, Saar, Cody, Vanderburg, Law, Kraus, Latham, Baranec, Riddle, Ziegler, Lund, Torres, Meibom, Aguirre, \& Wright}]{Curtis2020}
Curtis, J.~L., Ag{\"u}eros, M.~A., Matt, S.~P., {et~al.} 2020, When {{Do Stalled Stars Resume Spinning Down}}? {{Advancing Gyrochronology}} with {{Ruprecht}} 147, Vol. 904 ({IOP Publishing})

\bibitem[{Davidson(2014)}]{davidson2014}
Davidson, P. 2014, Geophysical Journal International, 198, 1832

\bibitem[{{Davidson}(2013)}]{2013GeoJI.195...67D}
{Davidson}, P.~A. 2013, Geophysical Journal International, 195, 67

\bibitem[{Delorme {et~al.}(2022)Delorme, Durocher, Brun, \& Strugarek}]{delorme2022}
Delorme, M., Durocher, A., Brun, A.~S., \& Strugarek, A. 2022, in {{SF2A-2022}}: {{Proceedings}} of the {{Annual}} Meeting of the {{French Society}} of {{Astronomy}} and {{Astrophysics}}. {{Eds}}.: {{J}}. {{Richard}}, 209--213

\bibitem[{DeRosa {et~al.}(2012)DeRosa, Brun, \& Hoeksema}]{Derosa2012}
DeRosa, M.~L., Brun, A.~S., \& Hoeksema, J.~T. 2012, The Astrophysical Journal, 757, 96

\bibitem[{Do~Nascimento {et~al.}(2023)Do~Nascimento, Barnes, Saar, De~Mello, Hall, Anthony, De~Almeida, Velloso, Da~Costa, Petit, Strugarek, Wargelin, Castro, Strassmeier, \& Brun}]{donascimentoHalelikeCycleSolar2023}
Do~Nascimento, J.-D., Barnes, S.~A., Saar, S.~H., {et~al.} 2023, The Astrophysical Journal, 958, 57

\bibitem[{{do Nascimento} {et~al.}(2020){do Nascimento}, {de Almeida}, Velloso, Anthony, Barnes, Saar, Meibom, {da Costa}, Castro, Galarza, {Lorenzo-Oliveira}, Beck, \& Mel{\'e}ndez}]{DoNascimento2020}
{do Nascimento}, Jr., J.~D., {de Almeida}, L., Velloso, E.~N., {et~al.} 2020, The Astrophysical Journal, 898, 173

\bibitem[{Donati {et~al.}(2006)Donati, Howarth, Jardine, Petit, Catala, Landstreet, Bouret, Alecian, Barnes, Forveille, Paletou, \& Manset}]{Donati2006}
Donati, J.-F., Howarth, I.~D., Jardine, M.~M., {et~al.} 2006, Monthly Notices of the Royal Astronomical Society, 370, 629

\bibitem[{Donati {et~al.}(2023)Donati, Lehmann, Cristofari, Fouqu{\'e}, Moutou, Charpentier, {Ould-Elhkim}, Carmona, Delfosse, Artigau, Alencar, Cadieux, Arnold, Petit, Morin, Forveille, Cloutier, Doyon, H{\'e}brard, \& {the Collaboration SLS}}]{donatiMagneticFieldsRotation2023}
Donati, J.-F., Lehmann, L.~T., Cristofari, P.~I., {et~al.} 2023, Monthly Notices of the Royal Astronomical Society, 525, 2015

\bibitem[{Durney \& Latour(1977)}]{Durney1977a}
Durney, B.~R. \& Latour, J. 1977, Geophysical \& Astrophysical Fluid Dynamics, 9, 241

\bibitem[{{Emeriau-Viard} \& Brun(2017)}]{Emeriau-Viard2017}
{Emeriau-Viard}, C. \& Brun, A.~S. 2017, The Astrophysical Journal, 846, 8

\bibitem[{Finley {et~al.}(2024)Finley, Brun, Strugarek, \& Cameron}]{finleyHowWellDoes2024}
Finley, A.~J., Brun, S.~A., Strugarek, A., \& Cameron, R. 2024, How Well Does Surface Magnetism Represent Deep {{Sun-like}} Star Dynamo Action?

\bibitem[{Finley \& Matt(2017)}]{Finley2017}
Finley, A.~J. \& Matt, S.~P. 2017, The Astrophysical Journal, 845, 46

\bibitem[{Folsom {et~al.}(2018)Folsom, Bouvier, Petit, L{\`e}bre, Amard, Palacios, Morin, Donati, \& Vidotto}]{folsom2018}
Folsom, C.~P., Bouvier, J., Petit, P., {et~al.} 2018, Monthly Notices of the Royal Astronomical Society, 474, 4956

\bibitem[{Gallet \& P{\'e}tr{\'e}lis(2009)}]{Gallet2009}
Gallet, B. \& P{\'e}tr{\'e}lis, F. 2009, Physical Review E, 80, 35302

\bibitem[{Gastine {et~al.}(2014)Gastine, Yadav, Morin, Reiners, \& Wicht}]{Gastine2014b}
Gastine, T., Yadav, R.~K., Morin, J., Reiners, A., \& Wicht, J. 2014, Monthly Notices of the Royal Astronomical Society: Letters, 438, 76

\bibitem[{G{\"u}del(2007)}]{gudel2007}
G{\"u}del, M. 2007, Living Reviews in Solar Physics, 4, 3

\bibitem[{Guinan {et~al.}(2002)Guinan, Ribas, \& Harper}]{guinan2002}
Guinan, E.~F., Ribas, I., \& Harper, G.~M. 2002, in Continuing the {{Challenge}} of {{EUV Astronomy}}: {{Current Analysis}} and {{Prospects}} for the {{Future}}, Vol. 264, 139

\bibitem[{Hale {et~al.}(1919)Hale, Ellerman, Nicholson, \& Joy}]{haleMagneticPolaritySunSpots1919}
Hale, G.~E., Ellerman, F., Nicholson, S.~B., \& Joy, A.~H. 1919, The Astrophysical Journal, 49, 153

\bibitem[{Hall {et~al.}(2021)Hall, Davies, {van Saders}, Nielsen, Lund, Chaplin, Garc{\'i}a, Amard, Breimann, Khan, See, \& Tayar}]{Hall2021}
Hall, O.~J., Davies, G.~R., {van Saders}, J., {et~al.} 2021, Nature Astronomy [\eprint[arxiv]{2104.10919}]

\bibitem[{Hansen {et~al.}(1995)Hansen, Kawaler, \& Arnett}]{hansen1995}
Hansen, C.~J., Kawaler, S.~D., \& Arnett, D. 1995, Physics Today, 48, 94

\bibitem[{Hotta {et~al.}(2023)Hotta, Bekki, Gizon, Noraz, \& Rast}]{hottaDynamicsLargeScaleSolar2023}
Hotta, H., Bekki, Y., Gizon, L., Noraz, Q., \& Rast, M. 2023, Space Science Reviews, 219, 77

\bibitem[{{Johns-Krull} \& Valenti(2000)}]{johns-krull2000}
{Johns-Krull}, C.~M. \& Valenti, J.~A. 2000, in Stellar {{Clusters}} and {{Associations}}: {{Convection}}, {{Rotation}}, and {{Dynamos}}, Vol. 198, 371

\bibitem[{Johnstone {et~al.}(2021)Johnstone, Bartel, \& G{\"u}del}]{Johnstone2021}
Johnstone, C.~P., Bartel, M., \& G{\"u}del, M. 2021, Astronomy \& Astrophysics, 649, A96

\bibitem[{{Kaneko} {et~al.}(2022){Kaneko}, {Hotta}, {Toriumi}, \& {Kusano}}]{2022MNRAS.517.2775K}
{Kaneko}, T., {Hotta}, H., {Toriumi}, S., \& {Kusano}, K. 2022, MNRAS, 517, 2775

\bibitem[{K{\"a}pyl{\"a} {et~al.}(2023)K{\"a}pyl{\"a}, Browning, Brun, Guerrero, \& Warnecke}]{kapyla2023}
K{\"a}pyl{\"a}, P.~J., Browning, M.~K., Brun, A.~S., Guerrero, G., \& Warnecke, J. 2023, Simulations of Solar and Stellar Dynamos and Their Theoretical Interpretation

\bibitem[{Karak {et~al.}(2020)Karak, Tomar, \& Vashishth}]{Karak2020}
Karak, B.~B., Tomar, A., \& Vashishth, V. 2020, Monthly Notices of the Royal Astronomical Society, 491, 3155

\bibitem[{Landin {et~al.}(2010)Landin, Mendes, \& Vaz}]{Landin2010}
Landin, N.~R., Mendes, L. T.~S., \& Vaz, L. P.~R. 2010, Astronomy and Astrophysics, 510, A46

\bibitem[{Lehmann {et~al.}(2023)Lehmann, Donati, Fouque, Moutou, Bellotti, Delfosse, Petit, Carmona, Morin, Vidotto, \& {consortium}}]{lehmannSPIRouRevealsUnusually2023}
Lehmann, L.~T., Donati, J.-F., Fouque, P., {et~al.} 2023, {{SPIRou}} Reveals Unusually Strong Magnetic Fields of Slowly Rotating {{M}} Dwarfs

\bibitem[{{Lorenzo-Oliveira} {et~al.}(2019){Lorenzo-Oliveira}, Mel{\'e}ndez, Galarza, Ponte, dos Santos, Spina, Bedell, Ram{\'i}rez, Bean, \& Asplund}]{lorenzo-oliveira2019}
{Lorenzo-Oliveira}, D., Mel{\'e}ndez, J., Galarza, J.~Y., {et~al.} 2019, Monthly Notices of the Royal Astronomical Society: Letters, 485, L68

\bibitem[{{Lorenzo-Oliveira} {et~al.}(2020){Lorenzo-Oliveira}, Mel{\'e}ndez, Ponte, \& Galarza}]{lorenzo-oliveira2020}
{Lorenzo-Oliveira}, D., Mel{\'e}ndez, J., Ponte, G., \& Galarza, J.~Y. 2020, Monthly Notices of the Royal Astronomical Society, 495, L61

\bibitem[{{Lorenzo-Oliveira} {et~al.}(2016){Lorenzo-Oliveira}, Porto De~Mello, \& Schiavon}]{lorenzo-oliveira2016}
{Lorenzo-Oliveira}, D., Porto De~Mello, G.~F., \& Schiavon, R.~P. 2016, A\&A, 594, L3

\bibitem[{Marsden {et~al.}(2014)Marsden, Petit, Jeffers, Morin, Fares, Reiners, {do Nascimento}, Auri{\`e}re, Bouvier, Carter, Catala, Dintrans, Donati, Gastine, Jardine, {Konstantinova-Antova}, Lanoux, Ligni{\`e}res, Morgenthaler, {Ram{\`i}rez-V{\`e}lez}, Th{\'e}ado, \& Van~Grootel}]{Marsden2014}
Marsden, S.~C., Petit, P., Jeffers, S.~V., {et~al.} 2014, Monthly Notices of the Royal Astronomical Society, 444, 3517

\bibitem[{Mathur {et~al.}(2023)Mathur, Claytor, Santos, Garc{\'i}a, Amard, Bugnet, Corsaro, Bonanno, Breton, {Godoy-Rivera}, Pinsonneault, \& Van~Saders}]{mathurMagneticActivityEvolution2023}
Mathur, S., Claytor, Z.~R., Santos, {\^A}. R.~G., {et~al.} 2023, The Astrophysical Journal, 952, 131

\bibitem[{Matt {et~al.}(2011)Matt, Do~Cao, Brown, \& Brun}]{Matt2011}
Matt, S.~P., Do~Cao, O., Brown, B.~P., \& Brun, A.~S. 2011, Astronomische Nachrichten, 332, 897

\bibitem[{Metcalfe {et~al.}(2022)Metcalfe, Finley, Kochukhov, See, Ayres, Stassun, {van Saders}, Clark, {Godoy-Rivera}, Ilyin, Pinsonneault, Strassmeier, \& Petit}]{metcalfe2022}
Metcalfe, T.~S., Finley, A.~J., Kochukhov, O., {et~al.} 2022, The {{Origin}} of {{Weakened Magnetic Braking}} in {{Old Solar Analogs}}

\bibitem[{Miesch {et~al.}(2006)Miesch, Brun, \& Toomre}]{Miesch2006a}
Miesch, M.~S., Brun, A.~S., \& Toomre, J. 2006, The Astrophysical Journal, 641, 618

\bibitem[{Nelson {et~al.}(2013)Nelson, Brown, Brun, Miesch, \& Toomre}]{Nelson2013}
Nelson, N.~J., Brown, B.~P., Brun, A.~S., Miesch, M.~S., \& Toomre, J. 2013, The Astrophysical Journal, 762, 73

\bibitem[{Noraz {et~al.}(2022{\natexlab{a}})Noraz, Breton, Brun, Garc{\'i}a, Strugarek, Santos, Mathur, \& Amard}]{noraz2022a}
Noraz, Q., Breton, S.~N., Brun, A.~S., {et~al.} 2022{\natexlab{a}}, A\&A, 667, A50

\bibitem[{Noraz {et~al.}(2022{\natexlab{b}})Noraz, Brun, Strugarek, \& Depambour}]{Noraz2021}
Noraz, Q., Brun, A.~S., Strugarek, A., \& Depambour, G. 2022{\natexlab{b}}, Astronomy \& Astrophysics, 658, A144

\bibitem[{O'Mara {et~al.}(2016)O'Mara, Miesch, Featherstone, \& Augustson}]{omaraVelocityAmplitudesGlobal2016}
O'Mara, B., Miesch, M.~S., Featherstone, N.~A., \& Augustson, K.~C. 2016, Advances in Space Research, 58, 1475

\bibitem[{Pedlosky(1987)}]{Pedlosky1987}
Pedlosky, J. 1987, Geophysical {{Fluid Dynamics}} ({Springer})

\bibitem[{Perri {et~al.}(2021)Perri, Brun, Strugarek, \& R{\'e}ville}]{Perri2021}
Perri, B., Brun, A.~S., Strugarek, A., \& R{\'e}ville, V. 2021, accept\'e dans ApJ [\eprint[arxiv]{2102.01416}]

\bibitem[{Petit {et~al.}(2008)Petit, Dintrans, Solanki, Donati, Auri{\`e}re, Ligni{\`e}res, Morin, Paletou, Ramirez~Velez, Catala, \& Fares}]{Petit2008}
Petit, P., Dintrans, B., Solanki, S.~K., {et~al.} 2008, Monthly Notices of the Royal Astronomical Society, 388, 80

\bibitem[{Reiners {et~al.}(2022)Reiners, Shulyak, K{\"a}pyl{\"a}, Ribas, Nagel, Zechmeister, Caballero, Shan, Fuhrmeister, Quirrenbach, Amado, Montes, Jeffers, Azzaro, B{\'e}jar, Chaturvedi, Henning, K{\"u}rster, \& Pall{\'e}}]{reiners2022}
Reiners, A., Shulyak, D., K{\"a}pyl{\"a}, P.~J., {et~al.} 2022, Magnetism, Rotation, and Nonthermal Emission in Cool Stars -- {{Average}} Magnetic Field Measurements in 292 {{M}} Dwarfs

\bibitem[{R{\'e}ville {et~al.}(2015)R{\'e}ville, Brun, Matt, Strugarek, \& Pinto}]{Reville2015a}
R{\'e}ville, V., Brun, A.~S., Matt, S.~P., Strugarek, A., \& Pinto, R.~F. 2015, The Astrophysical Journal, 798, 116

\bibitem[{Ribas {et~al.}(2005)Ribas, Guinan, G{\"u}del, \& Audard}]{Ribas2005}
Ribas, I., Guinan, E.~F., G{\"u}del, M., \& Audard, M. 2005, The Astrophysical Journal, 622, 680

\bibitem[{Rieutord(1987)}]{Rieutord1987}
Rieutord, M. 1987, Geophysical and Astrophysical Fluid Dynamics, 39, 163

\bibitem[{Santos {et~al.}(2021)Santos, Breton, Mathur, \& Garc{\'i}a}]{Santos2021}
Santos, A. R.~G., Breton, S.~N., Mathur, S., \& Garc{\'i}a, R.~A. 2021, The Astrophysical Journal Supplement Series, 255, 17

\bibitem[{Schaeffer(2013)}]{shtns}
Schaeffer, N. 2013, Geochemistry, Geophysics, Geosystems, 14, 751

\bibitem[{See {et~al.}(2015)See, Jardine, Vidotto, Donati, Folsom, Boro~Saikia, Bouvier, Fares, Gregory, Hussain, Jeffers, Marsden, Morin, Moutou, Do~Nascimento, Petit, Ros{\'e}n, \& Waite}]{See2015}
See, V., Jardine, M., Vidotto, A.~A., {et~al.} 2015, Monthly Notices of the Royal Astronomical Society, 453, 4301

\bibitem[{See {et~al.}(2019{\natexlab{a}})See, Matt, Finley, Folsom, Saikia, Donati, Fares, H{\'e}brard, Jardine, Jeffers, Marsden, Mengel, Morin, Petit, Vidotto, Waite, \& {and the BCool Collaboration}}]{See2019b}
See, V., Matt, S.~P., Finley, A.~J., {et~al.} 2019{\natexlab{a}}, The Astrophysical Journal, 886, 120

\bibitem[{See {et~al.}(2019{\natexlab{b}})See, Matt, Folsom, Saikia, Donati, Fares, Finley, H{\'e}brard, Jardine, Jeffers, Lehmann, Marsden, Mengel, Morin, Petit, Vidotto, \& Waite}]{See2019}
See, V., Matt, S.~P., Folsom, C.~P., {et~al.} 2019{\natexlab{b}}, The Astrophysical Journal, 876, 118

\bibitem[{Skumanich(1972)}]{Skumanich1972}
Skumanich, A. 1972, Astrophysical Journal, 171, 565

\bibitem[{Strugarek {et~al.}(2018)Strugarek, Beaudoin, Charbonneau, \& Brun}]{Strugarek2018}
Strugarek, A., Beaudoin, P., Charbonneau, P., \& Brun, A.~S. 2018, The Astrophysical Journal, 863, 35

\bibitem[{Strugarek {et~al.}(2017)Strugarek, Beaudoin, Charbonneau, Brun, \& {do Nascimento}}]{Strugarek2017}
Strugarek, A., Beaudoin, P., Charbonneau, P., Brun, A.~S., \& {do Nascimento}, J.-D. 2017, Science, 357, 185

\bibitem[{Takehiro {et~al.}(2020)Takehiro, Brun, \& Yamada}]{Takehiro2020}
Takehiro, S.-i., Brun, A.~S., \& Yamada, M. 2020, The Astrophysical Journal, 893, 83

\bibitem[{{van Saders} {et~al.}(2016){van Saders}, Ceillier, Metcalfe, Aguirre, Pinsonneault, Garcia, Mathur, \& Davies}]{VanSaders2016}
{van Saders}, J.~L., Ceillier, T., Metcalfe, T.~S., {et~al.} 2016, Nature, 529, 181

\bibitem[{Vidotto {et~al.}(2016)Vidotto, Donati, Jardine, See, Petit, Boisse, Boro~Saikia, H{\'e}brard, Jeffers, Marsden, \& Morin}]{Vidotto2016}
Vidotto, A.~A., Donati, J.-F., Jardine, M., {et~al.} 2016, Monthly Notices of the Royal Astronomical Society: Letters, 455, L52

\bibitem[{Vidotto {et~al.}(2014)Vidotto, Gregory, Jardine, Donati, Petit, Morin, Folsom, Bouvier, Cameron, Hussain, Marsden, Waite, Fares, Jeffers, \& {do Nascimento}}]{Vidotto2014a}
Vidotto, A.~A., Gregory, S.~G., Jardine, M., {et~al.} 2014, Monthly Notices of the Royal Astronomical Society, 441, 2361

\bibitem[{Viviani {et~al.}(2018)Viviani, Warnecke, K{\"a}pyl{\"a}, K{\"a}pyl{\"a}, Olspert, {Cole-Kodikara}, Lehtinen, \& Brandenburg}]{vivianiTransitionAxiNonaxisymmetric2018}
Viviani, M., Warnecke, J., K{\"a}pyl{\"a}, M.~J., {et~al.} 2018, Astronomy \& Astrophysics, 616, A160

\bibitem[{Warnecke {et~al.}(2016)Warnecke, K{\"a}pyl{\"a}, K{\"a}pyl{\"a}, \& Brandenburg}]{Warnecke2016}
Warnecke, J., K{\"a}pyl{\"a}, P.~J., K{\"a}pyl{\"a}, M.~J., \& Brandenburg, A. 2016, Astronomy and Astrophysics, 596, 1

\end{thebibliography}

%
%

\begin{appendix}

\section{Magnetic field decomposition definitions}\label{sec:appA}
In order to compare our simulations with the published results and figures of \citet{See2015,See2019,See2019b}, we used the formalism of \cite{Donati2006,Vidotto2016,folsom2018} (see also \citealt{Rieutord1987}). In this formalism, we made use of the vectorial spherical harmonics basis defined as 
\begin{equation}
  \label{eq:RST}
  \left\{
  \begin{array}{lcl}
    \Rlm{} &=& \Ylm{} \er \\
    \Slm{} &=& \gradperp \Ylm{} = \dth\Ylm{}\etheta + \frac{1}{\sin{\theta}}\dphi\Ylm{}\ephi\\
    \Tlm{} &=& \gradperp\times\Rlm{} = \frac{1}{\sin{\theta}}\dphi\Ylm{}\etheta  -\dth\Ylm{}\ephi
  \end{array}
  \right.\, ,
\end{equation}
where $Y^m_l$ are the orthonormalised classical spherical harmonics (defined as $\int Y^{m_1}_{l_1} \left(Y^{m_2}_{l_2}\right)^* {\rm d}\Omega = \delta_{l,l_1}\delta_{m_1,m_2}$). The vector spherical harmonics basis \ref{eq:RST} can be used to decompose any vector, ${\bf B}$, such that
\begin{equation}
    \label{eq:Bdecomp}
    {\bf B} = \sum\limits_{\substack{\ell,m\\\ell\leq \ell_{\rm cut}\\ -\ell\leq m\leq \ell}} \mathcal{A}_{l,m} \Rlm{} + \mathcal{B}_{l,m} \Slm{} + \mathcal{C}_{l,m} \Tlm{}\, .  
\end{equation}
Based on this decomposition and for a direct comparison with the observational results, we used the following definitions of the various magnetic field components in the present paper,
\begin{equation}
B_I^2 = B_{\rm rms}^2 = \frac{1}{4\pi} \sum\limits_{\substack{\ell,m\\\ell\leq \ell_{\rm max}\\ -\ell\leq m\leq \ell}} |\mathcal{A}_{\ell,m}|^2 + \ell(\ell+1) (|\mathcal{B}_{\ell,m}|^2 + |\mathcal{C}_{\ell,m}|^2)  \, ,
\end{equation}
\begin{equation}
    B_V^2 = \frac{1}{4\pi} \sum\limits_{\substack{\ell,m\\\ell\leq \ell_{\rm cut}\\ -\ell\leq m\leq \ell}} |\mathcal{A}_{\ell,m}|^2 + \ell(\ell+1) (|\mathcal{B}_{\ell,m}|^2 + |\mathcal{C}_{\ell,m}|^2) \, ,
\end{equation}
\begin{equation}
    B_{\rm pol}^2 = \frac{1}{4\pi} \sum\limits_{\substack{\ell,m\\\ell\leq \ell_{\rm cut}\\ -\ell\leq m\leq \ell}} |\mathcal{A}_{\ell,m}|^2 + \ell(\ell+1) |\mathcal{B}_{\ell,m}|^2 \, ,
\end{equation}
\begin{equation}
    B_{\rm \{dip;quad;oct\}}^2 = \frac{1}{4\pi} \sum\limits_{\substack{\ell,m\\\ell=\{1;2;3\}\\ -\ell\leq m\leq \ell}} |\mathcal{A}_{\ell,m}|^2 + \ell(\ell+1) |\mathcal{B}_{\ell,m}|^2 \, ,
\end{equation}
\begin{equation}
    B_{\rm tor}^2 = \frac{1}{4\pi} \sum\limits_{\substack{\ell,m\\\ell\leq \ell_{\rm cut}\\ -\ell\leq m\leq \ell}} \ell(\ell+1) |\mathcal{C}_{\ell,m}|^2  \, ,
\end{equation}
where $\ell_{\rm cut} = 5$ and  $\ell_{\rm max} = 2N_\theta /3$. The numerical module, SHTns \citep{shtns}, was used here to compute the different decompositions.  

\section{Relation between the stellar and the fluid Rossby number}\label{sec:appB}

\begin{figure}[ht!]
    \begin{center}
        \includegraphics[width=\linewidth]{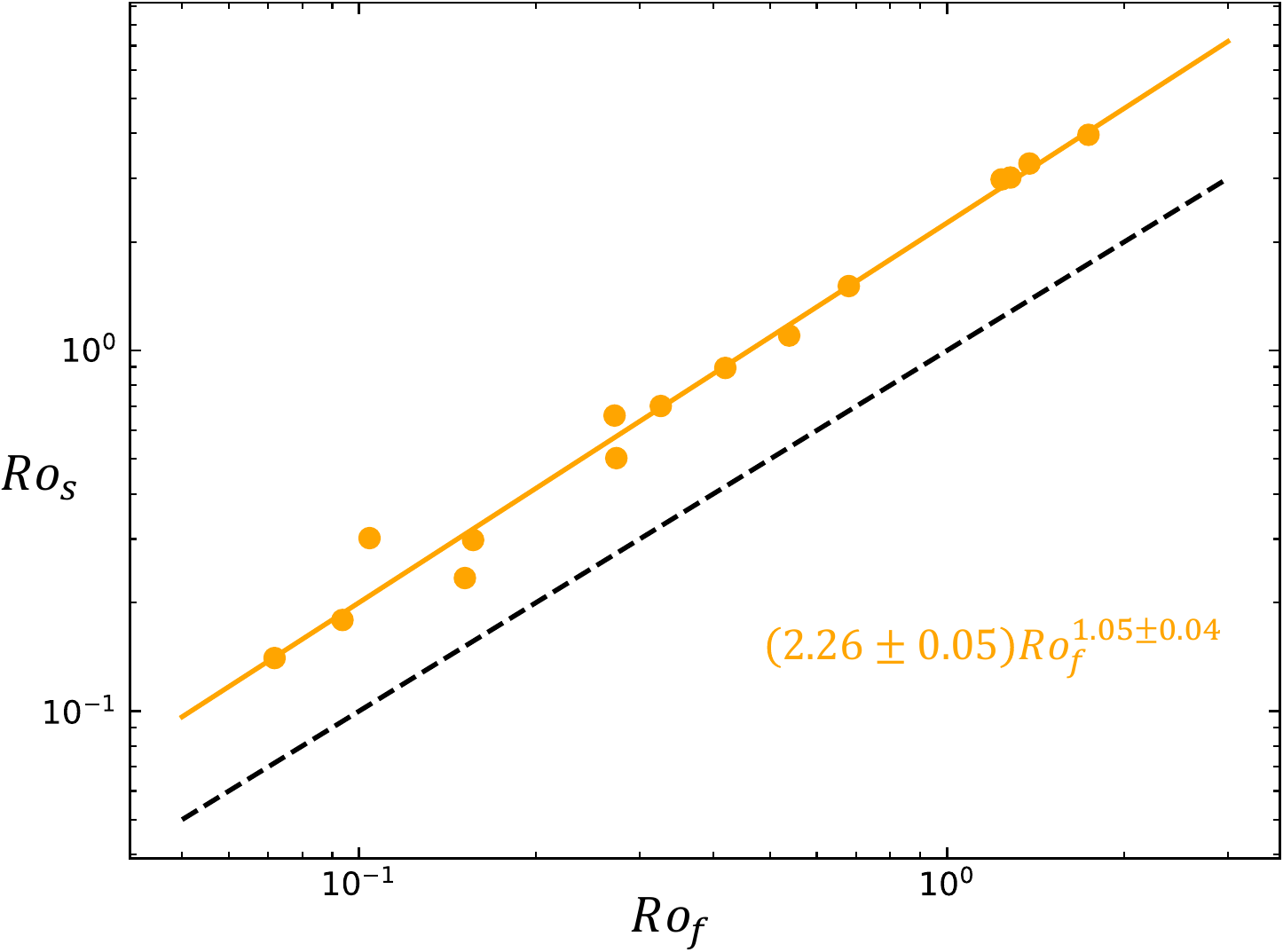}
    \end{center}
    \caption{Comparison of the stellar Rossby number, $Ro_{\rm s}$, as a function of the fluid one, $Ro_{\rm f}$, in the different models (orange dots) from \cite{Brun2022}. A linear regression is proposed with the solid orange line ($Ro_{\rm s} = (2.26\pm 0.05)Ro_{\rm f}^{1.05\pm 0.04}$), along with the direct proportionality, $Ro_{\rm s} = Ro_{\rm f}$, illustrated with the dashed black line.}\label{fig:9}
\end{figure}

In the present paper, we decided to adopt the ‘fluid’ definition of the Rossby number, $Ro_{\rm f}$, as it is the direct quantification of the advection term over the Coriolis term in the momentum equation of fluid dynamics, which we used in a previous study \cite{Brun2022} to characterise the different rotational and magnetic states of the set of simulations presented in Sect.~\ref{sec:sect2}. However, this number is not directly accessible from observations (see \citealt{noraz2022a}), and the observational trends we refer to in the present paper were computed with the ‘stellar’ definition of the Rossby number, $Ro_{\rm s}$ \citep{See2015,See2019,See2019b}. In order to compare our simulations fairly to the observations, we illustrate in Fig.~\ref{fig:9} the relationship between both definitions in our set of models.

The fluid Rossby number, $Ro_{\rm f}$, was computed following Eq.~\ref{eq:01} and taken in the middle of the convection zone. The stellar Rossby number, $Ro_{rm s}=P_{\rm rot}/\tau^{\rm CS}_c$, was computed similarly to \cite{See2015,See2019}, considering the empirical expression of the convective turnover time derived by \cite{cranmerTESTINGPREDICTIVETHEORETICAL2011}, which is $\tau^{\rm CS}_c=314.24\exp [ -\frac{T_{\rm eff}}{1952.5\,{\rm K}} - (\frac{T_{\rm eff}}{6250\,{\rm K}})^{18} ]+0.002$ days. We see that the relationship between both definitions is close to being linear, with a proportionality factor, $Ro_{\rm s}\simeq 2.26 Ro_{\rm f}$. An obvious second factor comes from their respective definitions, making the fluid Rossby number smaller, the remainder coming from small numerical differences. We use this calibration factor when plotting the observational trends as a function of the fluid Rossby number, $Ro_{\rm f}$, in Figs.~\ref{fig:3}, \ref{fig:4}, \ref{fig:5} and \ref{fig:6}.

\end{appendix}


\end{document}